\documentclass[12pt,preprint]{aastex}

\newcommand{\Ha}{H$\alpha$}
\newcommand{\OI}{[\ion{O}{1}]}
\newcommand{\NII}{[\ion{N}{2}]}
\newcommand{\SII}{[\ion{S}{2}]}
\newcommand{\OIII}{[\ion{O}{3}]}
\newcommand{\Hb}{H$\beta$}
\newcommand{\HeI}{\ion{He}{1}}
\newcommand{\HII}{\ion{H}{2}}

\shorttitle{Diffuse Ionized Gas in M33}
\shortauthors{Voges and Walterbos}

\begin{document}
\title{Detection of \OI $\:\lambda$6300 and Other Diagnostic Emission Lines \\ in the Diffuse Ionized Gas of M33 with Gemini-North}

\author{E. S. Voges and R. A. M. Walterbos}
\affil{Astronomy Department, New Mexico State University,
    Las Cruces, NM 88003; \email{esgerken@nmsu.edu, rwalterb@nmsu.edu}}
\newpage

\begin{abstract}
We present spectroscopic observations of diffuse ionized gas (DIG) in
M33 near the \HII$\:$ region NGC 604. We present the first detection
of \OI $\:\lambda$6300$\:$ in the DIG of M33, one of the critical
lines for distinguishing photo- from shock ionization models. We
measure \OI/\Ha$\:$ in the range of 0.04 to 0.10 and an increase in
this ratio with decreasing emission measure.  Our measurements of
\SII/\Ha$\:$ and \NII/\Ha$\:$ also rise with decreasing emission
measure, while our \OIII/\Hb$\:$ measurements remain fairly constant.
We have one tentative detection of \HeI$\:$ in the region of brightest
emission measure, with a ratio of \HeI/\Ha$=0.033\pm 0.019$,
indicating that the helium is at least partially ionized.  We compare
our observed emission line ratios to photoionization models and find
that field star ionization models do not fit our data well.  Leaky
\HII$\:$ region models are consistent with our data, without the need
to invoke additional ionization mechanisms to fit our \OI$\:$ or
\OIII$\:$ measurements.  The closest large \HII$\:$ region is NGC 604
and is therefore a likely candidate for the source of the ionizing
photons for the gas in this region.
\end{abstract} 

\section{Introduction}
The influence of massive stars and their deposition of energy into the
interstellar medium (ISM) can be seen through their creation of \HII$\:$
regions, superbubbles, filaments, and chimneys.  The full effect of
massive stars' presence in galaxies is still unknown, but they are
thought to also be responsible for the diffuse ionized gas (DIG, or
warm ionized medium, WIM) that has been observed in both the Milky Way
and external galaxies (e.g. \citealt{rey1991}; \citealt{rand1992};
\citealt{green1998}; \citealt{hoopes2003}).  In the Milky Way, the
diffuse ionized medium is known as the Reynolds layer.  This gas is
warm (8,000-10,000 K), has an electron density of 0.1-0.2 cm$^{-3}$,
and is the most massive component of the ionized Galactic ISM.
Studies of external galaxies have shown the DIG in other galaxies to
be similar in spatial extent to that of the Reynolds layer.

The Wisconsin H$\alpha$ Mapper (WHAM) Northern Sky Survey has revealed
that the ionized gas in the Milky Way is detectable in almost every
direction in the northern sky (\citealt{haff2003}).  The total power
required to ionize the local DIG in the Galaxy is about $4.0 \times
10^{-5}$ ergs s$^{-1}$ cm$^{-2}$, if all of the gas is assumed to be
at a temperature of $10^4$ K (\citealt{rey1990}).  Only OB stars can
meet or exceed this ionizing power requirement.  Studies of external
galaxies have suggested the power requirement from a galaxy's OB stars
is around 40\% to ionize the DIG (\citealt{hoopes1996};
\citealt{ferg1996}; \citealt{greenphd}), based on the observed DIG's
contribution to the galaxy's \Ha$\:$ luminosity.

One way to test whether the power source of the DIG is photoionization
by massive stars is to observe its optical spectrum.  Numerous
spectroscopic studies of the DIG in galaxies have revealed that DIG
has enhanced \SII$\:\lambda 6716$/H$\alpha$ and \NII$\:\lambda
6584$/H$\alpha$ compared to \HII$\:$ regions (e.g. \citealt{ferg1996};
\citealt{hoopes2003}).  These emission line ratios are well replicated
by pure photoionization models.  However, there are other emission
line ratios that are difficult to replicate and challenge pure
photoionization models.  Observations of edge-on galaxies have found
that the \OIII/H$\beta$ ratio may rise with distance from the plane
(\citealt{rand1998}) at a level that photoionization models may not be
able to reproduce.  Other mechanisms suggested to create this rising
\OIII/H$\beta$ ratio include shock heating (\citealt{shull79}) and
turbulent mixing layers (\citealt{slavin93}).  These models have some
support from observations that the \OIII$\:$ spectral line can be
broader than other spectral lines (\citealt{wang1997}).  The
\OI$\:\lambda$6300/6363 doublet is another indicator of the relative
contribution of shocks to the ionization of the DIG
(e.g. \citealt{dopita76}).  These lines are difficult to observe in
nearby galaxies due to the bright \OI$\:$ airglow line.  Little
information has yet been gleaned from this important emission line in
face-on galaxies.  Finally, little is known about the ionization
fraction of helium versus hydrogen, which can be probed by the ratio
of \HeI$\:\lambda 5876$ to \Ha.  Due to many bright skylines near
\HeI, there is insufficient reliable data available on \HeI$\:$
emission in face-on galaxies.  Only O7 stars or earlier produce the
high-energy photons required to ionize helium, so the presence of
\HeI$\:$ emission can give also clues as to the type of stars
responsible for the ionization of the DIG.

M33 is an ideal subject for spectroscopy of DIG because it is nearby
(0.84 Mpc away) and is suitably inclined.  M33 is blueshifted by -180
km s$^{-1}$ with respect to heliocentric velocities, making the
observation of \OI$\:$ possible.  To take full advantage of this small
blueshift, we used Gemini's nod-and-shuffle capabilities to get the
best sky subtraction possible.

\section{Observations and Data Reduction}
The data were taken using the Gemini Multi-Object Spectrograph (GMOS)
on the Gemini-North telescope by Gemini staff in the queue mode during
2005 January and February (as part of Program GN-2004B-Q-18).  We
obtained long-slit spectra with a slit 108'' long and 1'' wide of one
position in M33 (see Figure 1).  The target area is a region with an
\Ha$\:$ emission measure of about 25 pc cm$^{-6}$ near NGC 604.  We
were concerned with obtaining accurate sky line subtraction, so we
traded on-target time for frequent nodding off-target to get sky
exposures.  Three observing blocks of 3600 seconds open-shutter time
were obtained.  Each exposure had 15 nod-and-shuffle cycles (a
complete cycle being one exposure of the science target followed by
one exposure of the sky), with each cycle being 240 seconds long.
This came to 1800 seconds on-target for each observing block.  Our
nods were 9.2' long, taken in the direction parallel to the slit.

The R831 grating was used with the three mosaiced EEV 2048 $\times$
4068 CCDs, and the data were binned on chip by 2 in the wavelength
dimension and by 4 in the spatial dimension.  The resulting dispersion
scale was 0.68 $\mathrm{\AA}$ per pixel and the resulting spatial pixel scale
was 1.19 parsecs per pixel.  The spectral resolution measured on the arc
lamp lines was 3.0 \AA.  This spectral resolution was lower than
anticipated due to a focus problem.  The grating was tilted to a
central wavelength of 5858 \AA, with coverage from about 4850 to 6860
\AA.
 
The data reduction was carried out with the Gemini IRAF package.
After bias subtraction and multiplication by the gain for both the
science and sky spectra, the sky spectra were subtracted from their
corresponding science spectra.  The sky-subtracted spectra were then
combined, mosaiced, and trimmed to include only the middle third of
the CCD (i.e. where the science exposure was located).  The combined
spectrum was then flat-fielded, wavelength and flux calibrated, and
corrected for atmospheric extinction.  Standard star observations were
included in our program, however, they were not observed in
conjunction with our science observations.  As a result, we do not
have a good absolute flux calibration.  However, we do have a good
relative flux calibration, which is sufficient to compare emission
line strengths.

Charge-traps in the GMOS CCDs were evident in our spectra.  These
charge-traps are visible as streaks in the horizontal, or wavelength
dispersion direction.  \citet{abraham2004} suspect that these traps
are due to small detector defects that are repeatedly ``pumped'' by
the shuffle-and-pause action associated with nod-and-shuffle
observations.  Although the presence of these streaks is visually
unappealing, we did extensive noise analysis of the CCDs both near to
and far from these streaks, and the noise characteristics seemed
unaffected by the charge-traps.  We did, however, find that the
overall rms noise for the middle of the three CCDs was about 1.6 times
higher than the other two.  This was unfortunate for us as the
faintest of our desired emission lines (He I $\lambda$5876 and \NII
$\:\lambda$5755) fell on this CCD.  The combination of high overheads
(and thus reduced observation times), the intrinsic faintness of the
lines, and a noisy CCD made it impossible for us to detect \NII
$\:\lambda$5755, and we detected \HeI$\:$ in only one aperture.

\section{Emission Line Ratios}

We present results on our detection of \Ha, \SII $\:\lambda 6716$,
\NII $\:\lambda 6583$, \OI $\:\lambda 6300$, \OIII $\:\lambda 5007$,
and \Hb, as well as a tentative detection of \HeI$\:\lambda 5876$.  We
were unable to detect \NII $\:\lambda$5755 or \OI$\:\lambda 6363$.  We
detected \SII $\:\lambda 6731$, but the ratio between the two \SII$\:$
lines is constant in low density gas, so we do not present any results
for \SII $\:\lambda 6731$.  The emission line ratios presented here
were determined by averaging the spectrum along the spatial direction
over 42 pixels.  At an assumed distance of 0.84 Mpc to M33, this
corresponds to apertures approximately 50 pc in length.  The H$\alpha$
emission measure profile of the gas along the slit is fairly smooth,
so we chose equal sized apertures that increased our signal to noise
while preserving some spatial information along the slit.  To find the
uncertainties in the measurements, we multiplied the average rms noise
on either side of each emission line by the square root of the number
of pixels contained in the line.  The individual uncertainties were
added in quadrature for the uncertainty in the ratios.

\subsection{The \SII, \NII, \OIII, and \HeI$\:$ Emission Lines}
Figure 2 shows the ratios of \SII $\:\lambda 6716$/\Ha, \NII
$\:\lambda 6583$/\Ha, \SII/\NII, \OI $\:\lambda 6300$/\Ha, \OIII
$\:\lambda 5007$/\Ha, and \OIII/\Hb, along with the underlying
H$\alpha$ emission measure profile.  Given the location of the slit,
the x-axes on these plots can be taken not only as distance along the
slit, but also as increasing distance away from NGC 604.  In general,
the H$\alpha$ emission measure is seen to decrease with distance along
the slit.  If we are to assume that the density of the gas is about
constant along the slit and that NGC 604 is a contributing source of
ionizing photons in the region, then for a sufficiently thick disk of
gas we might expect a roughly spherical region of gas centered on NGC
604 to be influenced by its ionizing photons.  If that is the case,
then a decrease in the \Ha$\:$ emission measure is expected due to a
changing path length from the perspective of the observer as we move
away from NGC 604.  We note that given the distance of the slit to NGC
604, the gas disk would have to be thick enough to accommodate a
sphere of 820 pc.  Ultimately, this sort of geometry cannot be
distinguished from other possibilities, such as the increase in the
emission measure being due to a local feature like a filament.  The
ratio \SII/\Ha$\:$ appears to increase with decreasing emission
measure, and \NII/\Ha$\:$ exhibits a slight increase with decreasing
emission measure.  The ratio of \SII/\NII$\:$ does not vary a great
deal along the slit.  This behavior of \SII$\:$ and \NII$\:$ with
emission measure and their overall elevated values with respect to
\HII$\:$ regions is characteristic of DIG in other galaxies
(e.g. \citealt{rand1998} and \citealt{hoopes2003}).

We were able to measure \OI$\:$ in three of our seven apertures, and
we calculated upper limits for the other four apertures.  Our results
indicate an increase of \OI/\Ha$\:$ with decreasing emission measure.
We measured \OIII/\Ha$\:$ in in every aperture, and the ratio stays
fairly constant with emission measure.  We were able to measure
\OIII/\Hb$\:$ in the first three apertures.  The \Hb$\:$ emission line
fell near to the edge of the grating, where we had little sensitivity,
and thus we were unable to detect \Hb$\:$ in the regions of fainter
emission.  Our observed \Ha/\Hb$\:$ ratio in the first three apertures
is 3.1.  Given the uncertainty in our measurements, this value is
close enough to the intrinsic (unreddened) value of \Ha/\Hb$=2.86$, so
we did not do a correction for interstellar extinction.  We used our
observed \Ha/\Hb$\:$ ratio to infer the \Hb$\:$ intensities in the
last four apertures.  No correction was made for stellar absorption of
the \Hb$\:$ emission.  \citet{hoopes2003} found this correction in the
DIG of M33 to be about 10\%, but this will vary depending on the
contribution of the \Hb$\:$ emission in the DIG relative to the
\Hb$\:$ absorption by the stars.  Therefore, our measured
\OIII/\Hb$\:$ ratios might be too high by 10\%, or possibly more.
We had one tentative detection of \HeI$\:$ in the first aperture with
\HeI/\Ha$=0.033\pm0.019$.  With the large uncertainty associated with
this measurement, our detection does not improve upon the one in this
region given by \citet{hoopes2003}, so we will not go into any further
analysis of this result.

\subsection{\OI$\:$ Emission in the DIG}

Figure 3 shows our detection of \OI $\:\lambda 6300$ in the first
three apertures.  The continuum levels adopted for the flux
measurements are indicated on the spectra, and the noise was measured
locally.  This is the first detection of \OI $\:\lambda 6300$ in the
DIG of M 33, or any non-edge-on spiral other than the Milky Way.  Our
observed ratios of \OI/\Ha$\:$ are 0.038, 0.040, and 0.097 for
apertures 1, 2, and 3 respectively.  These measurements were made in
regions with emission measures of 48, 33, and 27 pc cm$^{-6}$.  For
comparison, the WHAM facility measured \OI/\Ha$\:$ in the warm ionized
Galactic ISM in three directions \citep{rey1998}.  Two of their
observations sampled DIG in the Galactic midplane, and the third at
$z=-300$ pc.  The observations sampled emission measures in the range
of 6 to 24 pc cm$^{-6}$ and they found \OI/\Ha$=0.012-0.044$.  The
detection of \OI$\:$ emission has implications regarding the relative
contribution of shock ionization, so we will come back to this
emission line in our comparison to photoionization models.

\section{Comparison with Photoionization Models}

\citet{hoopes2003} provide two CLOUDY photoionization
models for M33: the ``standard'' DIG model and the ``leaky \HII$\:$
region'' model.  The standard model simulates ionization by field OB
stars by embedding OB stars in gas with the properties of DIG.  The
leaky \HII$\:$ region model is a combination of two models.  The first
model is of an \HII$\:$ region leaking a fraction of its ionizing
photons.  The ionizing spectrum of a star hardens as it passes through
an \HII$\:$ region, thus the spectrum that emerges from the \HII$\:$
region is harder than the original stellar spectrum.  As a result,
\HII$\:$ region models with less leakage emit harder ionizing
continua.  The leaky \HII$\:$ region model then uses this hardened
spectrum as the ionization source for a standard DIG model.  In both
models, they varied two parameters: the ionizing stellar temperature
$T_\star$, and $q$, a measure of the ratio of the density of ionizing
photons to the density of electrons.  They varied the values of $q$ to
bracket values suitable for DIG, which are thought to be about $\log q
= -3$ to $-4$.  They used \citet{kurucz1991} ATLAS line-blanketed LTE
stellar atmospheres for the ionizing continua and varied $T_\star$
from 30,000 to 50,000 K.  They used the abundances of He, O, N, and S
from measurements made in NGC 604 \citep{vilchez1988}.  For the other
elements, they used one-third the Orion Nebula abundances (the average
of the factors suitable for oxygen and nitrogen).

Figure 4 compares the standard and leaky \HII$\:$ region model
predictions with our observed line ratios.  The \OIII/\Hb$\:$ line
ratios in apertures 4 through 7 have been inferred using our
\OIII/Ha$\:$ measurements and our observed \Ha/\Hb$\:$ ratio of 3.1.
Two leaky \HII$\:$ region models are shown: one with 30\% leakage from
the region and one with 60\% leakage.  From this figure we find that,
in general, the leaky \HII$\:$ region model with 30\% leakage best
fits our data.  This conclusion is most clear in the middle and bottom
plots in Figure 4, where the 30\% leakage model with $\log q=-3$ to
$-4$ brackets our observations and we have nearly constant inferred
stellar temperatures of 40,000 to 42,000 K.  In these two plots, our
observed line ratios push the standard and 60\% leakage models into
the regime where $\log q=-2$, which is thought to be too large a value
for $\log q$ in the DIG, and the inferred stellar temperatures are
higher than those inferred by the 30\% leakage model.  In the top
plot, at first glance we have several models that could be consistent
with our observations.  The hardened continuum that escapes from a
leaky \HII$\:$ region produces higher \NII/\Ha, \SII/\Ha, and
\OIII/\Hb$\:$ ratios than the standard model.  The 30\% leakage models
predict higher ratios for a given $q$ and $T_\star$ than the 60\%
leakage models.  These differences are what ultimately make the 30\%
leakage model most consistent with our data in the top plot as well,
where the increased \NII/\Ha$\:$ and \SII/\Ha$\:$ measurements in
apertures 6 and 7 cannot be reproduced by the standard or 60\% leakage
models without extremely high stellar temperatures.  The stellar
temperature inferred by the 30\% leakage model of about 42,000 K is
reasonable for ionizing stars within an \HII$\:$ region.  Our \HeI$\:$
detection implies the presence of an O7 or earlier star, which would
have a stellar temperature of about 40,000 K \citep{smith2002}.  We
also note that 30\% leakage from \HII$\:$ regions is consistent with
the study of the contribution of field OB stars to the ionization of
the DIG given by \citet{hoopes2000}. Their study indicates that field
stars only partially account for the ionization of the DIG, and
that 25-30\% leakage from \HII$\:$ regions is necessary to account for
the remaining DIG.

Hoopes \& Walterbos (2003) did not include \OI$\:$ in their diagnostic
plots for leaky \HII$\:$ regions with multiple values for $\log q$,
but they did include predictions for \OI/\Ha$\:$ for $\log q = -3$ for
various \HII$\:$ region leakage rates.  For 30\% \HII$\:$ region
leakage, our observed \OI/\Ha$\:$ ratios infer stellar temperatures
between 38,000 and 42,000 K.  For 60\% \HII$\:$ region leakage, our
observed \OI/\Ha$\:$ ratios infer stellar temperatures between 30,000
and 44,000 K.  Neither leaky \HII$\:$ region model requires
unrealistic stellar temperatures to reproduce our observed \OI/\Ha$\:$
ratios.  Therefore, these observations, along with similar results for
our \OIII$\:$ observations, indicate that photoionization from leaky
\HII$\:$ regions can comfortably provide the heating necessary to
create the amount of \OI$\:$ and \OIII$\:$ emission that we detect.
In this case, the candidate leaky \HII$\:$ region is NGC 604, a large
\HII$\:$ region about 0.5 kpc away from the slit.  We have not
considered shock ionization models here, and our emission line ratios
may or may not be consistent with them.  However, on energetic
grounds, OB stars must be the dominant source for the ionization of
the DIG, and our observations are consistent with pure photoionization
models.

\acknowledgments Based on observations obtained at the Gemini
Observatory, which is operated by the Association of Universities for
Research in Astronomy, Inc., under a cooperative agreement with the
NSF on behalf of the Gemini partnership: the National Science
Foundation (United States), the Particle Physics and Astronomy
Research Council (United Kingdom), the National Research Council
(Canada), CONICYT (Chile), the Australian Research Council
(Australia), CNPq (Brazil) and CONICET (Argentina).

\clearpage
\begin{figure}
\epsscale{0.5}
\plotone{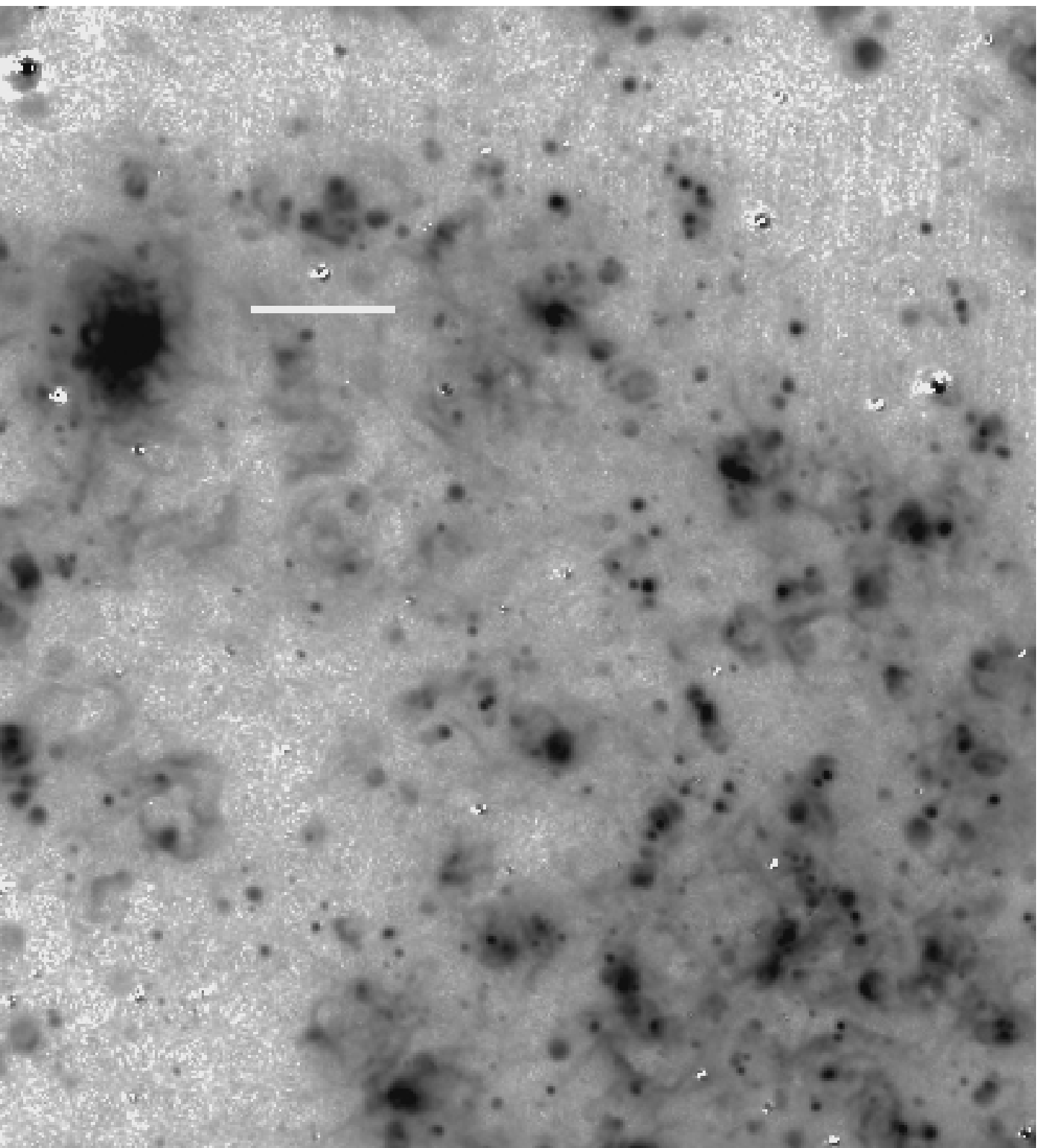}
\caption{\Ha $\:$image of M33 showing the slit location.  North is up
and east is left.  The slit is 108'' long, which at an assumed
distance to M33 of 0.84 Mpc, corresponds to 440 pc.  NGC 604 is the
large, bright \HII$\:$ region to the east of the slit.  The center of
NGC 604 is 820 pc away from the west end of the slit.}
\end{figure}

\clearpage
\begin{figure}
\epsscale{0.5}
\plotone{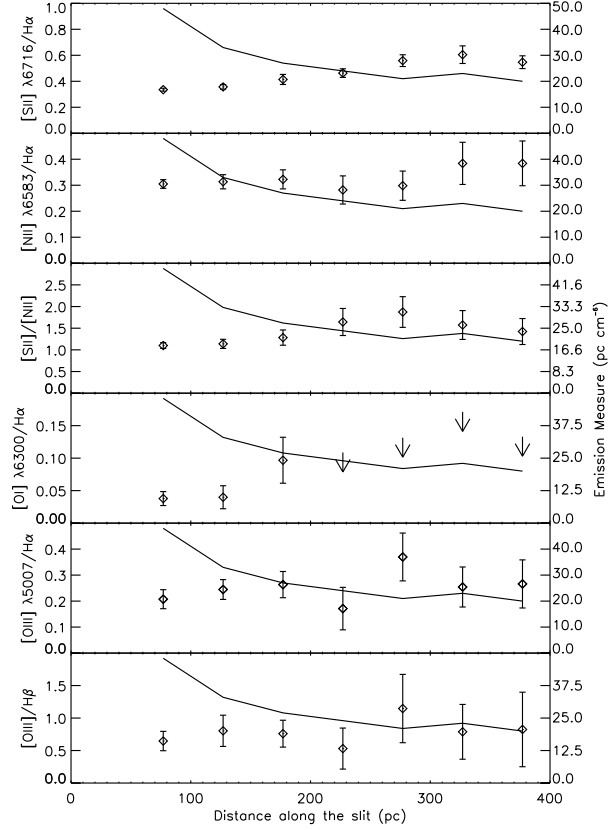}
\caption{Observed emission line ratios along the slit.  The first plot shows
\SII$\:\lambda 6716$/\Ha, the second plot shows \NII$\:\lambda
6583$/\Ha, the third shows \SII/\NII, the fourth shows \OI$\:\lambda
6300$/\Ha, the fifth shows \OIII$\:\lambda 5007$/\Ha, and the last
plot shows \OIII/\Hb.  The \Hb$\:$ fluxes in the last four apertures
were inferred from our \Ha$\:$ flux measurements, using our average
observed \Ha/\Hb$\:$ ratio of 3.1.}
\end{figure}

\clearpage
\begin{figure}
\epsscale{0.5}
\plotone{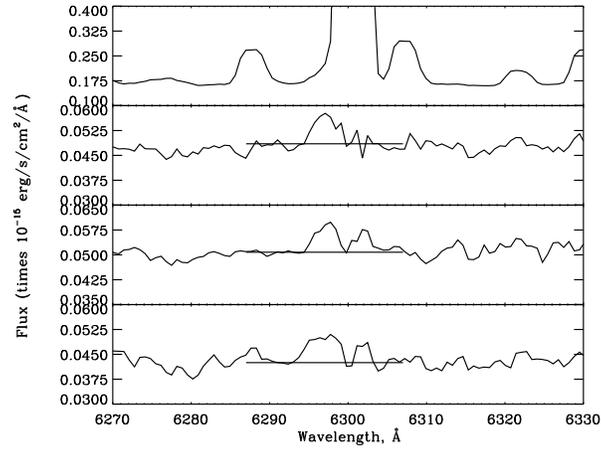}
\caption{Spectrum of apertures 1, 2, and 3 in M33 around \OI.  The top
plot shows the spectrum of aperture 1 before sky subtraction and the
bottom three plots, apertures 1, 2, and 3 after sky subtraction.  The
continuum level used for each measurement is indicated.  The feature
at 6297 \AA$\:$ is the \OI$\:$ emission from M33, and the feature to the
right of it is the residual from the \OI$\:$ airglow line subtraction.
Note that the scale of the plot before sky subtraction is 10 times
that of the plots after sky subtraction.}
\end{figure}

\clearpage
\begin{figure}
\epsscale{0.5}
\plotone{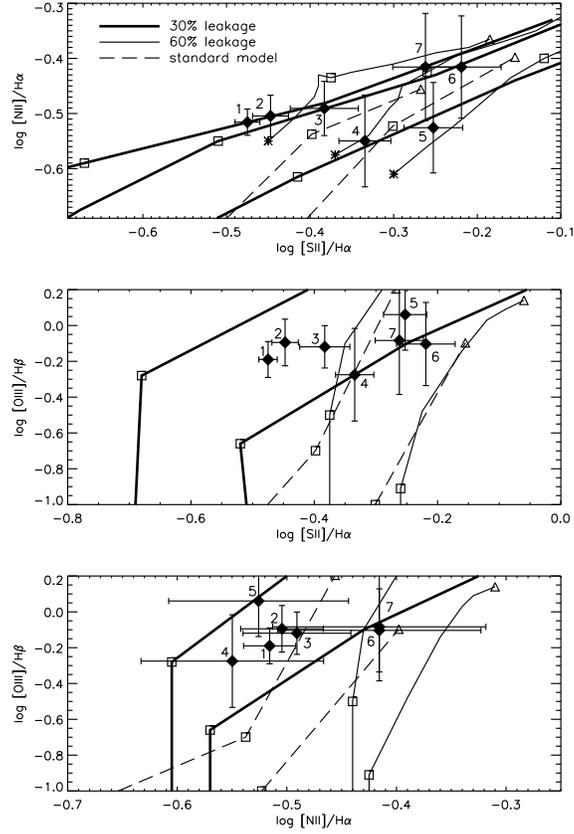}
\caption{A comparison of photoionization models to our observed line
ratios.  The thick solid lines show the predictions for the 30\%
\HII$\:$ region leakage models, the thin solid lines show the 60\%
leakage predictions, and the dashed lines shows the standard DIG model
predictions.  In the top plot we show three results for each model.
From right to left, the lines indicate values of $\log q =-5,-4,$ and
$-3$ respectively.  For the middle and bottom plot, from right to
left, we show results for $\log q=-4$ and $-3$.  The asterisks
indicate model stellar temperatures of 30,000 K; the squares, 40,000
K; and the triangles, 50,000 K.  Our observed line ratios are plotted
as solid diamonds.}
\end{figure}


\clearpage
\begin{thebibliography}{}
\bibitem[Abraham et al.(2004)]{abraham2004} Abraham, R.~G., et al.\ 
2004, \aj, 127, 2455
\bibitem[Dopita(1976)]{dopita76} Dopita, M.~A.\ 1976, \apj, 209, 
395 
\bibitem[Ferguson et al.(1996)]{ferg1996} Ferguson, A.~M.~N., 
Wyse, R.~F.~G., \& Gallagher, J.~S.\ 1996, \aj, 112, 2567 
\bibitem[Greenawalt(1998)]{greenphd} Greenawalt, B.~E.\ 1998, 
Ph.D.~Thesis,  
\bibitem[Greenawalt et al.(1998)]{green1998} Greenawalt, B., 
Walterbos, R.~A.~M., Thilker, D., \& Hoopes, C.~G.\ 1998, \apj, 506, 135 
\bibitem[Haffner et al.(2003)]{haff2003} Haffner, L.~M., 
Reynolds, R.~J., Tufte, S.~L., Madsen, G.~J., Jeans, K.~P., \& Percival, 
J.~W.\ 2003, \apjs, 149, 405
\bibitem[Hoopes et al.(1996)]{hoopes1996} Hoopes, C.~G., 
Walterbos, R.~A.~M., \& Greenwalt, B.~E.\ 1996, \aj, 112, 1429
\bibitem[Hoopes \& Walterbos(2000)]{hoopes2000} Hoopes, C.~G., \& 
Walterbos, R.~A.~M.\ 2000, \apj, 541, 597 
\bibitem[Hoopes \& Walterbos(2003)]{hoopes2003} Hoopes, C.~G., \& 
Walterbos, R.~A.~M.\ 2003, \apj, 586, 902 
\bibitem[Kurucz(1991)]{kurucz1991} Kurucz, R.~L.\ 1991, Precision 
Photometry:  Astrophysics of the Galaxy, 27 
\bibitem[Rand et al.(1992)]{rand1992} Rand, R.~J., Kulkarni, 
S.~R., \& Hester, J.~J.\ 1992, \apj, 396, 97 
\bibitem[Rand(1998)]{rand1998} Rand, R.~J.\ 1998, \apj, 501, 137 
\bibitem[Reynolds(1990)]{rey1990} Reynolds, R.~J.\ 1990, \apjl, 
349, L17
\bibitem[Reynolds(1991)]{rey1991} Reynolds, R.~J.\ 1991, IAU 
Symp.~144: The Interstellar Disk-Halo Connection in Galaxies, 144, 67 
\bibitem[Reynolds et al.(1998)]{rey1998} Reynolds, R.~J., 
Hausen, N.~R., Tufte, S.~L., \& Haffner, L.~M.\ 1998, \apjl, 494, L99 
\bibitem[Shull \& McKee(1979)]{shull79} Shull, J.~M., \& McKee, 
C.~F.\ 1979, \apj, 227, 131
\bibitem[Slavin et al.(1993)]{slavin93} Slavin, J.~D., Shull, 
J.~M., \& Begelman, M.~C.\ 1993, \apj, 407, 83 
\bibitem[Smith et al.(2002)]{smith2002} Smith, L.~J., Norris,
R.~P.~F., \& Crowther, P.~A.\ 2002, \mnras, 337, 1309
\bibitem[Vilchez et al.(1988)]{vilchez1988} Vilchez, J.~M., Pagel, 
B.~E.~J., Diaz, A.~I., Terlevich, E., \& Edmunds, M.~G.\ 1988, \mnras, 235, 
633 
\bibitem[Wang et al.(1997)]{wang1997} Wang, J., Heckman, T.~M., 
\& Lehnert, M.~D.\ 1997, \apj, 491, 114 
\end{thebibliography}
\end{document}